\documentclass{aa}
\usepackage{psfig}
\newcommand{\XMM}{\mbox{\it XMM-Newton\,}}
\newcommand{\xmm}{\mbox{\it XMM\,}}
\newcommand{\Chandra}{\mbox{\it Chandra\,}}

\newcommand{\ROSAT}{\mbox{\it ROSAT\,}}
\newcommand{\Einstein}{\mbox{\it Einstein\,}}

\newcommand{\Lcgs}{erg~s$^{-1}$}
\newcommand{\fcgs}{erg~s$^{-1}$~cm$^{-2}$}
\newcommand{\scinot}[2]{${#1}{\times}10^{{#2}}$}
\newcommand{\sciernot}[3]{$({#1}{\pm}{#2}){\times}10^{{#3}}$}
\newcommand{\mscinot}[2]{{{#1}{\times}10^{{#2}}}}

\newcommand{\radec}[7]{$\alpha={#1}^h{#2}^m{#3}\fs{#4}$, $\delta={#5}\degr{#6}\arcmin{#7}\arcsec$}

\begin{document}
\include{jrnl_macros}


   \title{The central region of M31 observed with \XMM 
   \thanks{Based on observations obtained with \XMM, an ESA science
   mission with instruments and contributions directly funded by ESA
   Member States and the USA (NASA).}}
   \subtitle{II. Variability of the individual sources.}
   \titlerunning{M31 observed with \XMM: II. Variability}

   \author{
J.P.~Osborne\inst{1} \and
K.N.~Borozdin\inst{2} \and
S.P.~Trudolyubov\inst{2} \and
W.C.~Priedhorsky\inst{2} \and
R.~Soria\inst{3} \and
R.~Shirey\inst{4} \and
C.~Hayter\inst{1} \and
N.~La~Palombara\inst{6} \and
K.~Mason\inst{3} \and
S.~Molendi\inst{6} \and
F.~Paerels\inst{7} \and
W.~Pietsch\inst{8} \and
A.M.~Read\inst{8} \and
A.~Tiengo\inst{5} \and
M.G.~Watson\inst{1} \and
R.G.~West\inst{1}
}

   \authorrunning{Osborne et al.}
	
   \offprints{K.~N.~Borozdin}
   \mail{kbor@lanl.gov}

   \institute{
Department of Physics \& Astronomy, University of Leicester, Leicester LE1 7RH, UK
\and
NIS Division, Los Alamos National Laboratory, Los Alamos, NM 87545, USA
\and
Mullard Space Science Laboratory, University College London, Holmbury St.~Mary, Dorking, RH5 6NT, UK
\and
Department of Physics, University of California, Santa Barbara, Santa Barbara, CA 93106, USA 
\and
XMM-Newton SOC, VILSPA-ESA, Apartado 50727, 28080 Madrid, Spain
\and
Istituto di Fisica Cosmica ``G.Occhialini'', Via Bassini 15, 20133, Milano, Italy
\and
Columbia Astrophysics Laboratory, Columbia University, New York, NY 10027, USA
\and
Max Planck Institut f\"{u}r Extraterrestrische Physik, Giessenbachstra{\ss}e, D-85741 Garching bei M\"{u}nchen, Germany
}

   \date{Received [date]; accepted [date]}

   \abstract{ We present the results of a study of the variability of
   X-ray sources in the central 30\arcmin\ of the nearby Andromeda
   Galaxy (M31) based on \XMM\ Performance Verification observations.
   Two observations of this field, with a total exposure time of about
   50~ks, were performed in June and December of 2000.  We found 116
   sources brighter than a limiting luminosity of
   \scinot{6}{35}~\Lcgs\ (0.3--12~keV, $d=760$~kpc). 
   For the $\sim$60 brightest sources, we searched for
   periodic and non-periodic variability; at least 15\% of
   these sources appear to be variable on a time scale of
   several months.  We discovered a new bright transient source
   $\sim$2.9\arcmin\ from the nucleus in the June observation; this
   source faded significantly and was no longer detected in December.
   The behaviour of the object is similar to 
   a handful of Galactic LMXB transients, most of which 
   are supposed to harbor black holes.
   We detected pulsations with a period of $\sim$865~s
   from a source with a supersoft spectrum.  The flux of this source
   decreased significantly between the two \xmm\ observations.  The
   detected period is unusually short and points to a rapidly spinning
   magnetized white dwarf.  The high luminosity and transient nature 
   of the source suggest its possible identification 
   with classical or symbiotic nova, some of which were observed
   earlier as supersoft sources.
}

   \maketitle
%

\section{Introduction}

\begin{figure*}
\vbox{\psfig{figure=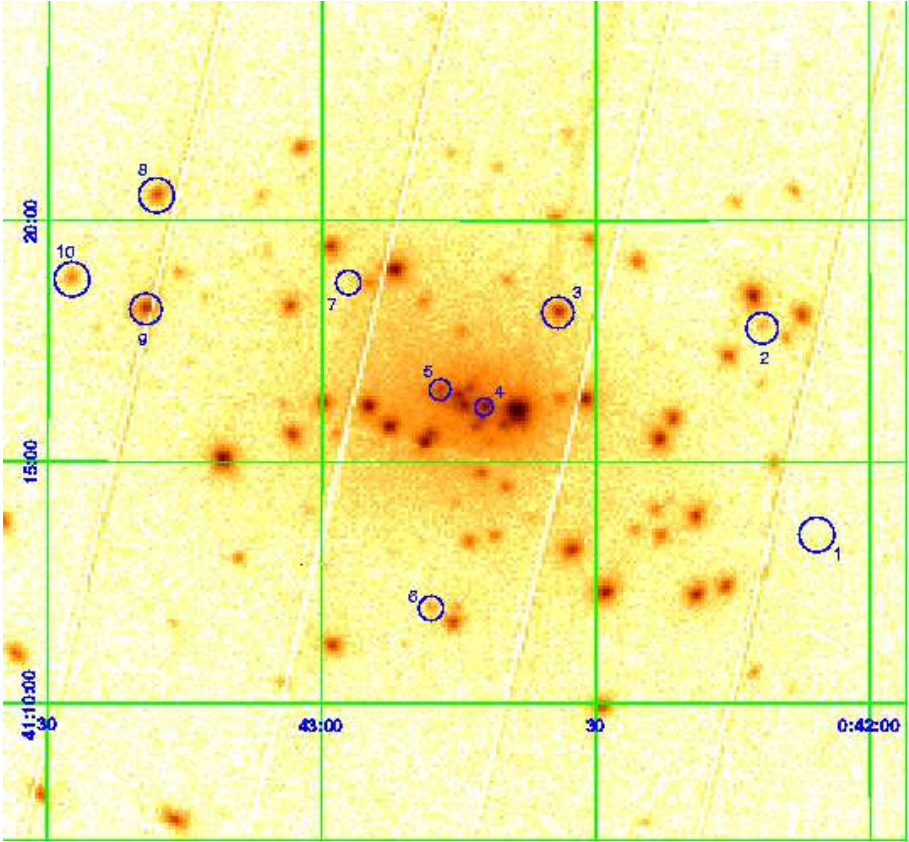,width=12cm}\vspace{-8.0cm}}
\hfill\parbox[b]{5.5cm}{
\caption{The 0.3 -- 10~keV EPIC PN image of the central region of M31.  
The data are from the 31-ks exposure taken on June 25, 2000. The
intensity of the color is proportional to the logarithm of the number of
counts collected (with a maximum of 5000 counts).  PN chip boundaries
are clearly visible. The positions of ten variable sources are
encircled and labelled with the numbers listed in Table~\ref{transients}.
\vspace{4.0cm} }
}
\label{fig:image}
\end{figure*}

At a distance of 760~kpc (\cite{vandenbergh00}; adopted throughout
this paper), M31 is close enough to allow detailed study of individual
sources within the galaxy using modern X-ray telescopes. Over 100
discrete X-ray sources in M31 were detected with the \Einstein\
observatory (Trinchieri \& Fabbiano 1991, hereafter \cite{tf};
\cite{vanspeybroeck79}).  Primini, Forman \& Jones (1993, hereafter
\cite{pfj}) reported on the detection of 86 X-ray sources in the
central 34\arcmin\ of M31 with the \ROSAT\ HRI\@.  Supper et~al.\ (2001,
hereafter \cite{su01}) recently published the complete catalog of 560
sources detected in a $\sim$10.7~deg$^2$ survey of M31 with the
\ROSAT/PSPC\footnote{This is an extension of the previously published
catalog of 396 sources (Supper et~al.\ 1997, \cite{su97}).}.
From the extrapolation of the luminosity distribution, PFJ concluded
that the detected population of X-ray sources could account for only
$\sim$15--26\% of the unresolved X-ray emission in M31, suggesting
that the remaining emission is truly diffuse or due to a new class of
X-ray sources.  Spectral analysis of \ROSAT\ data performed by Borozdin
\& Priedhorsky (2000\nocite{borozdin00}) and observations with \xmm\ 
(Shirey et~al.\ 2001, hereafter \cite{paperI}) and \Chandra\
(\cite{garcia01}) showed that the unresolved X-ray emission in the
bulge of M31 is significantly softer than most of the point sources
and can be approximated with an optically-thin thermal plasma model
($kT\sim$ 0.3~keV) as expected from truly diffuse emission.

\begin{table}
\caption[]{\label{observations}\XMM\ PV observations of the core of M31.}  
\begin{tabular}{llll}
\hline
\noalign{\smallskip}
Date \& Time (UTC)&Rev&Obs. ID&Exp.(ks)\\
\hline\hline
\noalign{\smallskip}
25/6/2000 (10:44--20:25)&100&0112570401&34.8$^a$/30.7$^b$\\ 
28/12/2000 (0:10--3:34)&193&0112570601&12.2$^a$/9.8$^b$\\ 
\hline
\end{tabular}
\begin{list}{}{}
\item[$^a$] for EPIC-MOS
\item[$^b$] for EPIC-PN
\end{list}
\end{table}

In the first \Chandra\ observation of M31, the nuclear source seen
with \Einstein\ and \ROSAT\ was resolved into five sources
(Garcia et al.\ 2000, hereafter \cite{G2000}).  
One of these sources is located within 1\arcsec\ of
the radio nucleus of M31 and exhibits an unusually soft X-ray
spectrum, suggesting that it may be associated with the central
super-massive black hole.  A few more pairs of previously unresolved
sources and a new transient were also detected within 30\arcsec\ of
the nucleus.

We report on observations of the Andromeda Galaxy (M31) carried
out with \XMM\ (\cite{jansen01}) during its Performance Verification
(PV) phase.  M31 was selected as an \xmm\ PV target in order to
demonstrate the capabilities of the mission in performing spectral and
timing studies in a field of point sources and extended emission.  In
\cite{paperI} we focused on the group properties of the X-ray point
sources and on the diffuse emission.  The spectral properties of
discrete X-ray sources in the \XMM\ exposures will be discussed in
Trudolyubov et~al.\ (2001, \cite{paperIII}).  In this present letter
we discuss the variability of individual X-ray sources in M31.  In
section~\ref{sect:obs}, we summarize the \XMM\ observations and our
data-reduction process.  In section~\ref{sect:srcs}, we discuss our
detections of transient and periodic variability in individual
sources. Finally, we present our conclusions.

\begin{table*}
\begin{center}
\caption[]{\label{transients}Bright variable X-ray sources detected in M31 with \XMM\ in 2000.}  
\begin{tabular}{clllllp{5.2cm}}
\hline
\noalign{\smallskip}
\# & Source name$^a$ & $\alpha$(2000) & $\delta$(2000) & F$_{Jun}^b$ & F$_{Dec}^b$ & Comments$^c$\\
\hline\hline
\noalign{\smallskip}
1 & XR~J004205.9+411329 & 00:42:05.9 & 41:13:29 & $<$2.7 & 12.6$\pm$2.6 & PFJ\#3\\ 
2 & XR~J004212.3+411800 & 00:42:12.3 & 41:18:00 & 1.8$\pm$1.3$^d$ & 17.5$\pm$4.1 & {\raggedright TF\#14, PFJ\#8}\\ 
3 & XMMU~J004234.1+411808 & 00:42:34.1 & 41:18:08 & 27.9$\pm$1.7 & $<$6.4 & \xmm\ X-ray nova (\S3.1)\\ 
4 & CXO~J004242.0+411608 & 00:42:42.2 & 41:16:09 & 58.1$\pm$2.1 & 45.6$\pm$3.0 & \Chandra\ transient (G2000 and \S3.2)\\
5 & RX~J0042.7+4116 & 00:42:47.2 & 41:16:28 & 9.5$\pm$1.3 & 94.0$\pm$3.6 & {\raggedright TF\#59, PFJ\#50, Su97\#198,\\ Su01\#195}\\ 
6 & XMMU~J004247.5+411158 & 00:42:47.5 & 41:11:58 & 4.0$\pm$1.3 & $<$3.0 &\\
7 & CXOU~J004257.1+411843 & 00:42:57.1 & 41:18:43 & $<$2.0 & 6.3$\pm$2.2 & seen on Oct 13, 1999 with \Chandra \\ 
8 & RX~J0043.3+4120 & 00:43:18.9 & 41:20:19 & 23.4$\pm$1.3$^e$ & 2.5$\pm$1.4$^e$ & SSS$^f$, TF\#87, Su97\#235, Su01\#235\\
9 & XMMU~J004319.4+411759 & 00:43:19.4 & 41:17:59 & 43.6$\pm$1.7$^{e,g}$ & $<$5.1$^{e,h}$ & SSS$^f$, 865-s pulsations (\S3.3)\\ 
10 & RX~J0043.4+4118 & 00:43:27.9 & 41:18:35 & 8.6$\pm$0.8$^e$ & 6.5$\pm$1.5$^e$ & {\raggedright SSS$^f$, TF\#89, PFJ\#80, Su97\#240, \\ Su01\#241, SNR}\\ 
\hline
\end{tabular}
\begin{list}{}{}
\item[$^a$] The names for $ROSAT$ and $Einstein$ sources 
were adopted from Su01. If the source had not been found in Su01, 
we used the acronym XR followed with truncated \xmm\ coordinates.
The sources seen first with \Chandra\ or \xmm\ are named
according to the naming conventions for these missions.
\item[$^b$] Source count rates for the \XMM\ observations in June 
and December of 2000 are given in units of $10^{-3}$counts/s (MOS1)
and are corrected for vignetting.  Quoted count rates are measured in the
0.3--10~keV energy range, except for supersoft sources (SSS), which
are measured in the 0.3--1.5~keV band.  In cases when sources are not
detected, 2$\sigma$ upper limits are listed.
\item[$^c$] References to the earlier observations: 
\cite{tf} = Trinchieri \& Fabbiano (1991);
\cite{pfj} = Primini, Forman \& Jones (1993);
\cite{su97}~=~Supper et~al.\ (1997);
\cite{su01}~=~Supper et~al.\ (2001);
\cite{G2000} = Garcia et~al.\ (2000).
\item[$^d$] Source near the chip edge; the detected count rate might be an underestimation of real count rate.
\item[$^e$] Source count rates for 0.3-1.5 keV band.
\item[$^f$] Supersoft source.
\item[$^g$] With contribution from a nearby faint source.
\item[$^h$] Upper limit is defined by the contribution of a nearby source.
\end{list}
\end{center}
\end{table*}


\section{\XMM\ Observations}\label{sect:obs}

The bulge of M31 was observed with \XMM\ on June 25, 2000 and again on
Dec 28, 2000 (see Table~\ref{observations}).  The observations were
centered on the core of M31 (\radec{00}{42}{43}{0}{+41}{15}{46.0}
J2000), with a field of view of 30\arcmin\ in diameter for the three
European Photon Imaging Camera (EPIC) instruments. The two EPIC MOS
instruments (\cite{turner01}) and the EPIC PN (\cite{struder01})
operated in full-window mode with the medium optical blocking filter.
The Optical/UV Monitor Telescope (OM; \cite{mason01}) filter wheel was
set to the blocked position during the June observation.  During the
December observation, exposures were obtained with the B and UVW1 filter
of the OM; results of these observations will be presented elsewhere.

A background flare occurred during the final 5~ks of the EPIC exposures
from the June observation.  Data obtained during this background flare
were excluded from the variability studies.

We used the \XMM\ Science Analysis System (versions 4.1 and 5.0.1) to
reduce the EPIC data to calibrated event lists, produce images, and
extract light curves.  A combination of SAS programs and external
software was applied to further analyze the data.  The data from MOS1,
MOS2 and PN were extensively compared to confirm the consistency of
our results.  The variability of individual sources within the longest
(June) observation was studied by Fourier analysis.  
The value of the pulsation period detected in one
of the supersoft sources (see below) was refined by the epoch-folding
method.


\section{Individual X-ray sources}\label{sect:srcs}

The EPIC images of the central 30\arcmin\ of M31 (Fig.~\ref{fig:image}
represents part of EPIC PN image) contain more than 100 discrete X-ray
sources as well as unresolved emission near the centre (see Paper I
for more details).
The detections include sources seen with \Einstein, \ROSAT, or
\Chandra\ (\cite{tf,pfj,su97,su01,G2000}) as well as new sources.  The
transient source discovered by \Chandra\ (\cite{G2000}) is clearly
seen in both \XMM\ observations (\S3.2).  In the June observation,
there is a bright new transient which was not seen previously (\S3.1). 
Several other bright sources demonstrated significant variability 
on half of a year time scale, so that uncertainties of count rate 
measurements do not overlap for the two observations.
We have analyzed 66 sources with limiting count rate\footnote{Some 
of the sufficiently bright sources were 
excluded from our analysis because of the confusion with nearby sources
or because they were too close to the chip edge} 
4 counts/ks and found 10 variables. 
These sources are listed in Table~\ref{transients} along with their \xmm\
positions, June and December count rates in MOS1, and any relevant 
identifications (though only count rates in MOS1 are cited, we selected
the sources based on the data from all three \XMM\ instruments).
The source positions were derived from comparison with \Chandra\
calibrated images of the central region of M31.  The estimated
positional error of 1\arcsec--3\arcsec\ (in diameter) depends on the
distance of the source from the boresight axis, its intensity and
nearby sources.  We estimate 3\arcsec\ as a conservative error limit
for the \xmm\ positions presented in this paper.

Taking into account the total number of sufficiently bright 
detected sources, we conclude that at least $\sim$15\% of 
all sources in the central field of M31 are variable on a time scale 
of several months.  We believe, however, that this value should 
be considered a lower limit, because the sensitivity 
of our analysis was a function of the source flux, and the variability 
of many of the fainter sources would not be detected.  
Below we discuss in more detail several individual
variable sources in M31.

\subsection{X-ray Nova in M31}

We found a new bright X-ray source in EPIC PN and EPIC MOS images from
the June observation.  The coordinates of the new source are presented
in Table~\ref{transients}.  The source was neither detected in
previous \Einstein\ and \ROSAT\ observations nor has been reported from
\Chandra\ observations.  During the June observations the flux from 
the source was \sciernot{1.57}{0.09}{-13}~\fcgs\ (0.3--10~keV), 
which corresponds to a luminosity of
$\sim\mscinot{1.1}{37}$~\Lcgs\ in the cited energy band 
assuming a power-law spectrum (see \cite{paperIII}).
The source flux faded to below the background level before
the next \XMM\ observation of the same field on Dec 28, 2000. 
The upper limit for the count rate for this second observation 
was less than 6.4~counts/ks (2$\sigma$ upper limit for MOS1).  
The detected luminosity of the source during the 
June 25 observation, as well as its several-month time 
scale to fade to quiescence, is typical for bright
X-ray transients in our Galaxy (see reviews by \cite{tsh96} and
\cite{chen97}).

\subsection{\Chandra\ transient}

A bright transient source designated CXO~J004242.0+411608 was discovered with
\Chandra\ during a 17.5~ks observation of the core of M31 on Oct 13, 1999
(\cite{G2000}).  We detected an X-ray source at the same location
during both our \XMM\ PV observations.  The flux of the source
measured on June 25, 2000 was equal to
\sciernot{5.5}{0.5}{-13}~\fcgs\ ($L_x\sim\mscinot{3.8}{37}$~\Lcgs).
The source was detected again on Dec 28, 2000 with flux of
\sciernot{4.2}{1.0}{-13}~\fcgs. Including both the \Chandra\ and
\xmm\ detections, the source has remained bright for more than 14
months, with only slight possible fading ($\sim$25\%) between the two
\XMM\ observations separated by half of a year.  The X-ray
luminosity and spectral shape detected by \XMM\ (\cite{paperIII})
correspond to those of the hard/low state of Galactic black hole
transients (e.g., \cite{tl95}).  Typically Galactic X-ray transients
fade significantly in less than $\sim$100 days (see \cite{chen97} for
a review); however, a long plateau in a hard spectral state was
observed from the X-ray transient GRS~1716-249
(\cite{sunyaev94,rev98}).  We should mention that our sampling does
not allow us to distinguish reliably between single and repeated
outbursts.  It will be interesting to follow the evolution of
CXO~J004242.0+411608 during future planned observations with \XMM.

\subsection{Supersoft 865-s pulsator}

\begin{figure}
\psfig{figure=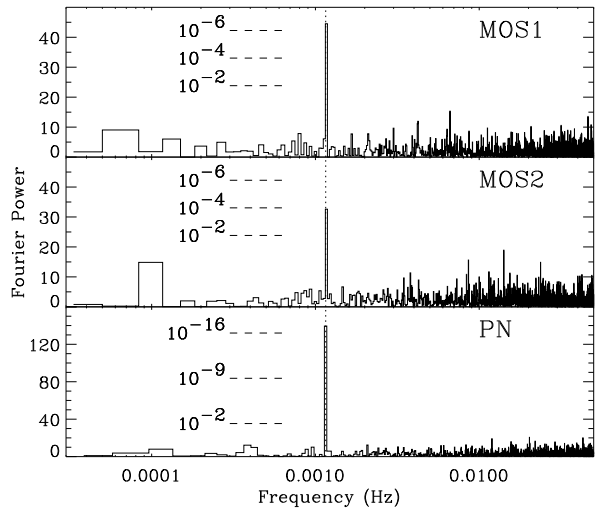,width=8.8cm,angle=0}\vspace{-4.3cm}
\caption{Power density spectra for a supersoft source in M31 obtained 
from each of the three EPIC instruments during the observation on June
25, 2000.  The peak power from each of the three instruments occurs at
a frequency corresponding to to a period of $\sim$865~s.  The vertical
dotted line denotes the best fit value of the period found by the
epoch-folding method.  Detection thresholds are indicated by
horizontal dashed lines labelled with the approximate probability
that any of the 1500 or 1300 frequency bins in each PDS (for MOSs and
PN respectively; no averaging applied) would have a noise value
exceeding the indicated power level (\cite{vaughan94}).}
\label{fig:pds}
\end{figure}

\begin{figure}
\psfig{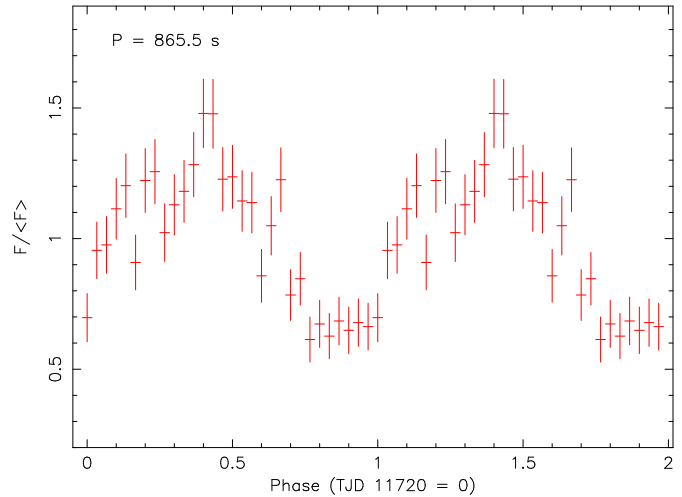}
\caption{Light curve of the pulsating supersoft source in M31, 
folded using a period of 865.5~s. The data is from the June 25, 2000
observation, and counts from all three EPIC instruments have been
used. Contribution of the background is within the error bars.}
\label{fig:lightcurve}
\end{figure}

We detected significant oscillations in the X-ray flux from one of the
bright variable sources in our field (source \#9 in
Table~\ref{transients}).  The Fourier power density spectra (PDS) from
each of the three independent EPIC instruments (Fig.~\ref{fig:pds})
show a highly significant periodic signal from this source during the
June observation, with a best-fit period of 865.5$\pm$0.5~s.  The
folded light curve for this source (Fig.~\ref{fig:lightcurve}) is
quasi-sinusoidal with an amplitude of about 40\%.

In June 2000 this 865-s pulsator had a spectrum similar to the three
sources in the \xmm\ field previously identified by Kahabka
(1999\nocite{kahabka99}) as supersoft source (SSS) candidates based on
their \ROSAT/PSPC hardness ratios and luminosity (one was identified
with a supernova remnant).  The temperature of the blackbody spectral 
fit was kT$_{bb}$=61$\pm$2 eV with reduced $\chi^2$=1.5 (for EPIC-MOS data).
All four SSS exhibit blackbody-like
spectra with effective temperatures in the range kT$\sim$50--150~eV
and no detectable X-ray emission above $\sim$1.5~keV (see
\cite{paperIII} for more detailed discussion of spectral parameters).  
Such sources are typically interpreted as accreting
white dwarfs in binary systems, powered by nuclear burning
of the accreted matter on their surfaces (see \cite{kah97} for a review).

The source count rate during the June~25 observation was equal to
44~counts/ks in the 0.3--1.5~keV energy range (MOS1, see
Table~\ref{transients}).  By Dec.~28 the source count rate had faded
down to a level below 5~counts/ks (MOS1); thus, it was not possible to
detect pulsations during the second observation.
All three supersoft sources in the NE portion of our field (sources
\#\# 8, 9 and 10) faded between the June and December observations
(see Table~\ref{transients}).  We are aware of a possible sensitivity
degradation in this region of MOS1; however, the result is confirmed
by MOS2 and PN data.  In particular, the count rate of the supersoft
pulsator as detected by MOS2 dropped down from 41$\pm$1.5~counts/ks in
June to 3$\pm$1.5~counts/ks in December, while many other sources in
the field showed no significant variability or became brighter in
December.

We note that there is a nearby source
(\radec{00}{43}{21}{2}{+41}{17}{52}, J2000) with a harder spectrum
which is much fainter than the supersoft pulsator in the June
observation but dominates this region of the sky for the December
observation.  The distance between two sources is about 10$\arcsec$,
so that they are spatially resolved with \XMM, but their counts are not
completely spatially separated on the detector.  Because 
the relative flux of the harder source
was much lower during the June observation, it did not significantly
affect our timing and spectral analysis of the SSS.  The \ROSAT/PSPC
would barely be able to resolve the two sources, so \ROSAT's
RX~J0043.3+4117 (\cite{su01}\#236) may include contributions from both
of them, although its position coincides with that of the harder
source rather than the SSS pulsator.

The 865.5-s period is the shortest among all known SSSs (\cite{gr2000}).
Interpreted as the binary orbital period, it would be too short to
accommodate a main-sequence companion and would suggest a degenerate
secondary.  It may be more plausible to assume that the pulsations
indicate that the white dwarf possesses a magnetic field large enough to
modulate the X-ray emission yet not so large that the spin and orbital
periods are locked, e.g., as in intermediate polars or DQ~Her stars
(see review by \cite{cordova95}).  However, the luminosity of the
object $\sim\mscinot{1.7}{37}$~\Lcgs\ (0.3--1.5~keV) is several orders
of magnitude higher then typical for intermediate polars 
luminosity range L$_X$=10$^{31}$--10$^{34}$ (see e.g. \cite{patt94}).
The high luminosity and transient nature of the pulsator 
may indicate steady burning in a post-nova stage, as has been 
observed in a few classical and symbiotic novae (\cite{kah97}).
An alternative explanation for the nature of the SSS could be
a double degenerate polar similar to RX~J1914.4+2456 (\cite{hm95,ram00}).
An interpretation of the source as a foreground object is
unlikely due to the lack of an optical counterpart in \xmm\ 
OM images up to the limiting magnitude of $\sim$19 in B filter.

\subsection{Variability of other sources within a single observation}

We have searched for coherent periodic modulation on time scales from
$\sim$10 to $\sim$1000~s for about 60 of the brightest sources in the
\XMM\ field of view; however, only in one case (see previous
subsection) was significant variability detected.  The 90\% upper
limits to periodic modulation fractions, calculated as the ratio of
the sine amplitude to the constant flux level for periods of 10\,000, 300,
and 10~s were obtained in all other cases.  These upper limits vary
from 3.6\% for the brightest source to $\sim$30\% for the faintest
sources of the sample.

The lack of detectable variability for many of the individual sources
in M31 may seem surprising compared with rich variability observed
from Galactic sources.  Our observations so far have included mainly
the bulge of M31, where low-mass X-ray binaries (LMXBs) are most
prevalent.  Such systems commonly show dips, bursts, and
quasi-periodic variability rather than coherent pulsations.  We expect
to detect more X-ray pulsars during planned \xmm\ observations of M31
in fields along the disk of M31, where population~I stars and
high-mass X-ray binaries (HMXBs) dominate.
We must note also that the sensitivity to variability depends strongly
on the brightness of the source, and hence our data are not very
sensitive to the variability of the many faint sources.  Finally,
we did not perform timing analysis for time scales shorter than $\sim$10~s
because the absolute time calibration was not possible with the Observation
Data Files available.

We have also looked for non-periodic variations on all accessible
timescales for the same set of objects.  No X-ray burst has been
detected, however our sensitivity to bursts is restricted to
a relatively small luminosity interval by the low count rate of 
most sources and by the Eddington limit.

\section{Conclusions}

In this letter we present results obtained from \XMM\ PV observations
of M31.  Two observations separated by half of a year were carried out
in June and December of 2000.  This letter is the second paper in the
series describing results of these observations.

Significant variability of individual sources was detected both
between the two \xmm\ observations and in comparison with earlier results
of other missions.  At least $\sim$15\% of the sources appear
to be variable, and we consider this value to be a conservative lower
limit.

A new bright transient source was detected during the observation of
June 25 but faded before the December observation. Probably it is an
LMXB transient source, similar to a handful of such sources observed
in our Galaxy, most of which are supposed to harbour black holes.

Another transient source, first detected by \Chandra\ (\cite{G2000}),
was bright during both \XMM\ observations.  The flux of the source did
not change significantly in observations separated by six months,
which is not typical for Galactic X-ray transients but is reminiscent
of the behavior of the black hole candidate GRS~1716-249 during its
1993--1994 outbursts (\cite{sunyaev94,rev98}).

X-ray pulsations with period of $\sim$865.5~s and quasi-sinusoidal
pulse profile were detected from one of the supersoft sources in our
field.  It was bright during the June observation, but had faded such
that it became undetectable in December.  The period of the detected
pulsations is the shortest among known SSSs.  A likely source of the
pulsations is a magnetized rapidly spinning white dwarf; however, the
luminosity of the source is much higher than for typical CV systems.
The detected X-ray flux may be generated by steady nuclear burning 
in a post-nova stage of a classical or symbiotic nova.

\begin{acknowledgements}
We thank all the members of the \XMM\ teams for their work building,
operating, and calibrating the powerful suite of instruments on-board.
We also thank the referee, Dr.\ T.\ Oosterbroek, for his helpful
comments.
\end{acknowledgements}

\end{document}